\documentclass[twocolumn,showpacs,preprintnumbers,amsmath,amssymb]{revtex4}
\usepackage{graphicx}
\usepackage{dcolumn}
\usepackage{bm}
\newcommand{\be}{\begin{equation}}
\newcommand{\ee}{\end{equation}}

\newcommand{\ber}{\begin{eqnarray}}
\newcommand{\eer}{\end{eqnarray}}

\newcommand{\bra}{\langle}
\newcommand{\ket}{\rangle}

\newcommand{\bs}[1]{\ensuremath{\boldsymbol{#1}}}
\newcommand{\spinmat}[4]{
     \left( \begin{array}{cc}
               #1 & #2  \\
               #3 & #4  
             \end{array}  \right) } 
\begin{document}

\title{Dynamical mean-field approximation for unitary Fermi gas}
\author{Nir Barnea}
\email{nir@phys.huji.ac.il}
\affiliation{The Racah Institute of Physics, The Hebrew University,
91904 Jerusalem, Israel.\\
     Institute for Nuclear Theory, University of Washington, 98195 Seattle,
  Washington, USA}
\date{\today}

\begin{abstract}
Dynamical mean-field approximation with explicit pairing is utilized to study
the properties of a two-component Fermi gas at unitarity.
The problem is approximated by the lattice Hubbard Hamiltonian,
and the continuum limit is realized by diluting the lattice.
We have found that at zero temperature  the predictions of this theory 
for the energy and the pairing gap agree
remarkably well with the results of full numerical Monte-Carlo simulations.
Investigating the evolution of the system with temperature we identify the
existence of 
a second order phase transition associated with a jump in the heat capacity
and the collapse of the pairing gap. 
\end{abstract}

\maketitle
{\it Introduction --}
The properties of a dilute Fermi gas with interparticle distance much larger
than the effective range $r_e$ depend on only two parameters.
The scattering length, $a_s$ which is sufficient to characterize the
interaction and interparticle distance or the Fermi momentum $k_F$. 
When the pair interaction is fine tuned to
create a two-body bound state with zero energy, the scattering length diverges
and the Fermi momentum remains the only length scale. 
At this point, commonly referred to as unitarity, the system acquires
universality, as its properties become independent of the nature of its  
constituents.
Dilute neutron gas where $a_s$ is an order of magnitude
larger than $r_e$ is a natural example for such system. 
Unitarity conditions can also be achieved with cold Fermi atoms
fine tuned near a Fesbach resonance \cite{Courtrille98}.

Regardless of the strength of the interaction, the zero temperature
ground state of a Fermi gas with attractive two-body force is a superconductor.
As the attraction increases the nature of the system changes
from a BCS superconductor at weak coupling to a gas of fermionic pairs forming
a  
Bose-Einstein condensation (BEC) at strong coupling. At unitarity the system
is in between these two limits and is neither a BCS superconductor nor a BEC. 

In this work we apply the dynamic mean field theory (DMFT), introduced
over a decade 
ago by Georges and Kotliar \cite{Georges92}, to study the properties 
of unitary Fermi gas. 
In this theory,
a lattice problem is mapped into a self-consistent embedded
impurity problem \cite{Georges96}. This mapping becomes exact in the limit of
infinite spatial dimensions $d\longrightarrow\infty$ due to the
localization of the self-energy 
$\Sigma(\bs{k},\omega)\rightarrow \Sigma(\omega)$ \cite{Metzner89}. For finite
dimensions DMFT is no longer exact, yet can be regarded as a useful 
approximation in which a purely local self-energy is assumed, hence
the name dynamical mean field approximation (DMFA). 

Applying the DMFA to study the properties of continuous Fermi gas, we
first construct a lattice version of
the problem \cite{Nir08} and then seek the continuum limit which
for finite gas densities corresponds to vanishing lattice filling.

{\it The lattice formulation --} 
The many-body Hamiltonian describing a dilute low energy Fermi gas
is, 
\begin{eqnarray} \label{H_continuum}
  H & = & -\frac{\hbar^2}{2 m}
      \sum_{\sigma}\int d\bs{x} \psi^{\dagger}_{\sigma}(\bs{x})
     \nabla^2 \psi_{\sigma}(\bs{x}) 
\cr & & 
    + \frac{1}{2}V_0 \sum_{\sigma} 
      \int d\bs{x} \psi^{\dagger}_{ \sigma}(\bs{x})
                   \psi^{\dagger}_{-\sigma}(\bs{x})
                   \psi^{}_{-\sigma}(\bs{x})
                   \psi^{}_{ \sigma}(\bs{x})\;,
\end{eqnarray}
where $\psi^{}_{ \sigma}(\bs{x})$ are the fermionic field operators.
To construct a lattice version 
of this continuum Hamiltonian we represent the configuration space as an $L^3$
cubic lattice,  
where $L$ is the number of sites in 
each spatial direction. The time direction is kept continuous.
Next, we replace the position and momentum variables by the grid indices
$
  \bs{x} \rightarrow  \bs{n} 
\;,\;
  \bs{p}\rightarrow \frac{2\pi}{L}\bs{k} \;,
$
where $\bs{n},\bs{k}$ are integer vectors. The grid position and momentum
are given by $a\bs{n}$ and $\bs{p}/a$, where $a$ is the
lattice spacing.
The fermionic fields 
$
  \psi_{\sigma}(\bs{x}) \rightarrow (a)^{-3/2} \psi_{\bs{n}\sigma}
$
are discretized 
to obey the anti-commutation relations
$
  \{\psi_{\bs{n}\sigma},\psi^{\dagger}_{\bs{n}\sigma}\}=
  \delta_{\sigma\sigma'}\delta_{\bs{n n'}}\,.
$
The corresponding lattice theory is the Hubbard Hamiltonian
\begin{equation}\label{H_lattice}
  H = -t \sum_{\sigma \bs{n}\bs{n}'} D_{\bs{n n}'} 
                       \psi^{\dagger}_{\bs{n} \sigma}
                       \psi_{\bs{n}' \sigma}
    +  U \sum_{\bs{n}} 
                   \psi^{\dagger}_{\bs{n}\, \uparrow}
                   \psi^{       }_{\bs{n}\, \uparrow}
                   \psi^{\dagger}_{\bs{n}\, \downarrow}
                   \psi^{       }_{\bs{n}\, \downarrow}\;,
\end{equation}
where $t=\frac{\hbar^2}{2 m a^2}$, and $U = V_0/a^3$. $D$ is the hopping
operator, 
$
 (D \psi_{\sigma})_{\bs{n}}=\sum_i 
(\psi_{\bs{n}+e_i,\sigma}-2\psi_{\bs{n},\sigma}+\psi_{\bs{n}-e_i,\sigma}),
$
where, $e_j$ is a unit vector in the direction $j$.
The spectra of the free lattice Hamiltonian is given by 
\begin{equation} 
\epsilon_{\bs{p}} = \frac{\hbar^2}{m a^2} D_{\bs{p}}
 \hspace{5mm} ; \hspace{5mm}
D_{\bs{p}} = 2 \sum_i \sin^2\frac{p_i}{2} \,.
\end{equation}
In the following we shall use natural units setting $\hbar=m=1$.
The strength of the two-body interaction can be related to the two--body
scattering length $a_s$ through 
summation of the ladder diagrams for two-fermions interacting at zero energy,
zero temperature and zero chemical potential, $\mu\rightarrow 0^-$,
\cite{Papenbrock99,Chen04}
\begin{equation}\label{def_as}
\frac{1}{4\pi a_s} = \frac{1}{V_0} + \frac{C}{2 a} 
\end{equation}
where
\begin{equation}\label{def_c}
 C = \int_{-\pi}^{\pi}\frac{d\bs{p}}{(2\pi)^3}\frac{1}{D_{\bs{p}}} \approx
 0.50532 \;.
\end{equation}
At unitarity $V_{0c}=-2a /C$, so $U_c=-7.91576 t$.
In contrast to the original continuum Hamiltonian (\ref{H_continuum}), the
lattice theory (\ref{H_lattice}) acquires an
effective range due to the finite lattice spacing.
The ratio between this effective range, 
$r_{eff} \approx 2 a/\pi^2$ \cite{Esbensen97}, and the average interparticle
distance is approximately $\frac{1}{2\pi}\sqrt[3]{ n }$,
where $ n $ is the average lattice filling, i.e. the number of
particles per site. For
finite lattice filling  
$ 0.1 \geq  n \geq 0.01 $ the corresponding average interparticle
distance is roughly $12$ to $24$ times the effective range.


{\it DMFA -} Integrating out the fermionic degrees of freedom on all lattice
sites but one - the impurity site - the DMFT maps
the many-body Hamiltonian (\ref{H_lattice}) into a single site effective action
determined self-consistently from a bath with which the impurity
site hybridizes. 
The impurity site can be taken to be a single lattice node \cite{Georges96} 
or a cluster of nodes \cite{Maier05-CDMFT}. Large cluster size improves
the accuracy of the DMFA. As the computation
complexity of the DMFA grows substantially with cluster size it is of great
interest to  asses
the quality of the simplest approximation within this framework, namely the
single node impurity.

Using the Nambu formalism, for a system with a superconducting long-range
order,  
the impurity site effective action takes the form
\cite{Georges96}
\begin{eqnarray}\label{S_eff}
 S_{eff} & = & -\int_0^{\beta}d\tau \int_0^{\beta}d\tau'
 \Psi^{\dagger}(\tau)\hat{\cal G}^{-1}_0(\tau-\tau')  
 \Psi^{}(\tau') 
\cr & & 
 - U \int_0^{\beta}d\tau \, 
  c^{\dagger}_{\uparrow  }(\tau)c^{}_{\uparrow  }(\tau)
  c^{\dagger}_{\downarrow}(\tau)c^{}_{\downarrow}(\tau) \;,
\end{eqnarray}
where $\beta=1/T$ is the inverse temperature, 
$\Psi^{\dagger} \equiv (c_{\uparrow}^{\dagger},c_{\downarrow})$
are the Nambu spinors, and $\hat{\cal G}_0(\tau)$ is given by
\begin{equation}
 \hat{\cal G}_0(\tau) = 
     \spinmat{{\cal G}_0(\tau)  }{ {\cal F}_0(\tau)}
             {{\cal F}^*_0(\tau)}{-{\cal G}_0(-\tau)}  \;.
\end{equation}
In the following we shall use the "hat" notation for the Nambu matrices.
The corresponding impurity Green's function is given by
\begin{eqnarray}\label{G_eff}
 \hat{\cal G}(\tau) & \equiv &
      -\bra T \Psi^{}_{i}(\tau) \Psi^{\dagger}_{i}(0) \ket_{S_{eff}} 
=
     \spinmat{{\cal G}(\tau)  }{ {\cal F}(\tau)}
             {{\cal F}^*(\tau)}{-{\cal G}(-\tau)} ,
\cr & &
\end{eqnarray}
where 
$
  {\cal G}(\tau) = -\bra T c^{}_{\sigma}(\tau) c^{\dagger}_{\sigma}(0)
                      \ket_{_{S_{eff}}} 
$
and
$
  {\cal F}(\tau) = -\bra T c^{}_{\uparrow}(\tau) c^{}_{\downarrow}(0)
                      \ket_{_{S_{eff}}} \;.
$

In the DMFA the interaction effects are taken into account through the
{self-energy} matrix
\begin{equation}
\hat \Sigma(i\omega_n) = \spinmat{\Sigma(i\omega_n)}{S(i\omega_n)}
                                  {S(i\omega_n)}{-\Sigma^*(i\omega_n)}
\end{equation}
deduced from the Dyson equation
\begin{equation}\label{self_energy}
 \hat \Sigma(i\omega_n)=\hat {\cal G}^{-1}_0(i\omega_n)
                       -\hat {\cal G}^{-1}(i\omega_n)\,,
\end{equation}
where $\omega_n={(2n+1)\pi}/{\beta}$ are the Matsubara frequencies.
Here and in the following, we have assumed that the symmetry of the pairing is
such that the off diagonal self energy obeys 
$S(i\omega_n)=S^*(-i\omega_n)$.
The connection to the physical {lattice} is made
through the self-consistency requirement that the impurity Green's function is
equal to the local lattice Green's function 
$ \hat {\cal G}(\tau)=\hat G(\tau)$, or
$ {{\cal G}(\tau)}={G_{}(\tau)}$, and ${{\cal F}(\tau)}={F}(\tau)$,
where 
\begin{eqnarray}
\hspace{-2mm}
  G_{} &=& \sum_{\bs{k}}
       \frac{-i\omega_n + \mu-\epsilon_{\bs{k}}-\Sigma^{*}(i\omega_n)}
            {|i\omega_n + \mu-\epsilon_{\bs{k}}-\Sigma(i\omega_n)|^2
             +S^{2}(i\omega_n)} \,,
\cr
\hspace{-2mm}
  F_{} &=&-\sum_{\bs{k}}
            \frac{S(i\omega_n)}
                 {|i\omega_n+\mu-\epsilon_{\bs{k}}-\Sigma(i\omega_n)|^2
                  +S^2(i\omega_n)} \,.
\end{eqnarray}

{\it Solving the Impurity Model -}
Caffarel and Krauth \cite{Caffarel94} proposed to approximate the effective
impurity action through an Anderson model 
\begin{eqnarray}\label{H_And}
{\cal H}_{And} & = & {\cal H}_0 + {\cal H}_I
   \cr
   & = & \sum_{l,\sigma} \tilde\epsilon_l a^{\dagger}_{l
  \sigma}a^{}_{l \sigma}    
+ \sum_{l,\sigma} \tilde V_l ( a^{\dagger}_{l \sigma} c_{\sigma}
  +c^{\dagger}_{\sigma} a^{}_{l \sigma} ) 
\cr & + &
 \sum_{l,\sigma} \tilde D_l ( a^{\dagger}_{l \sigma} c^{\dagger}_{-\sigma}
  +c^{}_{-\sigma} a^{}_{l \sigma} )
+ U n_{\uparrow} n_{\downarrow} \;,
\end{eqnarray}
where the interaction of the fermionic field
$c_{\sigma}$ with the auxiliary bath fermions $a_{l\sigma}$
generate both the normal and abnormal components of the ``free"
impurity Green's function ${\cal G}_0, {\cal F}_0$. 
This goal is achieved by choosing the parameters of the Anderson model 
$\tilde \epsilon_l, \tilde V_l, \tilde D_l$ 
to minimize the difference between the $\hat {\cal G}_0$ and 
$\hat {\cal G}_0^{And}$ over a finite range of frequencies 
$|\omega_n|\leq \omega_N$. 
The ``free'' Anderson's Green's function $\hat {\cal G}_0^{And}$
is calculated by numerical inversion of  ${\cal H}_0$.
When the number $n_s$ of fermionic fields, $a^{}_{l}$ and $c$, is smaller than
$6$ standard diagonalization methods can be used to solve 
${\cal H}_{And}$.
For $T=0$ the Lanczos method makes
a calculation with as many as $n_s=10$ fermionic fields feasible
\cite{Georges96}. 

{\it Extracting the Physics --}
Solving the DMFA equations 
yields the local approximation for the self-energy 
$\hat\Sigma(\bs{k},i\omega_n)\approx \hat\Sigma(i\omega_n)$, and
correspondingly 
\begin{equation}\label{G_ko}
  G(\bs{k},i\omega_n) =
       \frac{-i\omega_n + \mu-\epsilon_{\bs{k}}-\Sigma^{*}(i\omega_n)}
            {|i\omega_n + \mu-\epsilon_{\bs{k}}-\Sigma(i\omega_n)|^2
             +S^{2}(i\omega_n)} \,,
\end{equation}
and
\begin{equation}\label{F_ko}
  F(\bs{k},i\omega_n) =
             -\frac{S(i\omega_n)}
                 {|i\omega_n+\mu-\epsilon_{\bs{k}}-\Sigma(i\omega_n)|^2
                  +S^2(i\omega_n)} \,.
\end{equation}
Once $G$ and $F$ are known, thermodynamic quantities such as the number of
particles per site, the energy, and the superconducting gap can be 
calculated through the 
Matsubara sums, 
\begin{equation}\label{dens_per_site}
  n  = \frac{1}{\beta}\sum_{\sigma \bs{k}}
    \sum_{n=-\infty}^{\infty} e^{i 0^+} 
    G(\bs{k},i\omega_n) \,
\end{equation}
\begin{equation}\label{delta_0}
  \Delta_0 = \frac{U}{\beta}\sum_{\bs{k}}
    \sum_{n=-\infty}^{\infty} e^{i 0^+} 
    F(\bs{k},i\omega_n) \,
\end{equation}
and
\begin{equation}
  E  = \frac{1}{2}\frac{1}{\beta}\sum_{\sigma \bs{k}} 
                \sum_{n=-\infty}^{\infty} e^{i 0^+}
                (i\omega_n+\epsilon_{\bs{k}}+\mu) G(\bs{k},i\omega_n)\;.
\end{equation}
The number of particles $n$ and the gap $\Delta_0$ can  
also be calculated directly from the Anderson's Hamiltonian
(\ref{H_And}). If $\hat {\cal G}=\hat G$ these results would coincide. However,
at best $\hat {\cal G}\approx \hat G$, so this is not always the case.
Since the Matsubara sum contains explicit dependence on the lattice
density of 
state it provides a more reliable estimate for $n$ and $\Delta_0$. 
We shall use the relative difference 
$\delta_n =(n_M-n_{And})/n_M$ between the
Matsubara density $n_M$ and the Anderson's model density $n_{And}$ 
as a measure for the quality of the impurity solution.
For very accurate solution of
the impurity model we expect that $\delta_n \ll 1$.
When $\delta_n$ deviates substantially from
zero it is an indication that the number of auxiliary fields we have used in
the solution of (\ref{S_eff}) is not sufficient.

Direct evaluation of the Matsubara sums is impractical. Having calculated
$\Sigma(i\omega_n),S(i\omega_n)$ over a finite range of frequencies we
evaluate the thermodynamic observables in the following manner. 
We set $\Sigma_{\infty}=Re(\Sigma(i\omega_N))$, and
$S_{\infty}=S(i\omega_N)$, 
assuming
that at $\omega_N$, the largest frequency we explicitly consider, the
self-energy has already acquired its asymptotic value. 
Using these quantities we construct the 
Green's function components
 $G_{\infty}, F_{\infty}$ by the appropriate
substitutions of $\Sigma_{\infty}, S_{\infty}$ in
Eqs. (\ref{G_ko}), (\ref{F_ko}). 
The Matsubara sums with $G_{\infty}, F_{\infty}$ are evaluated analytically 
to obtain $n_{\infty}, E_{\infty}$ etc.
Then for the frequencies in the range $|\omega_n|\leq \omega_N$
we calculate the difference between $G$ and $G_{\infty}$ to obtain 
\begin{equation}
  n = n_{\infty} + \frac{1}{\beta}\sum_{\sigma \bs{k}} 
                \sum_{n=-{N}}^{{N}} 
                \left( G(\bs{k},i\omega_n)-G_{\infty}(\bs{k},i\omega_n)
                \right)\;,
\end{equation}
and equivalently for any other
thermodynamic observable of interest.

\begin{figure}[ht]
\includegraphics[height=6cm]{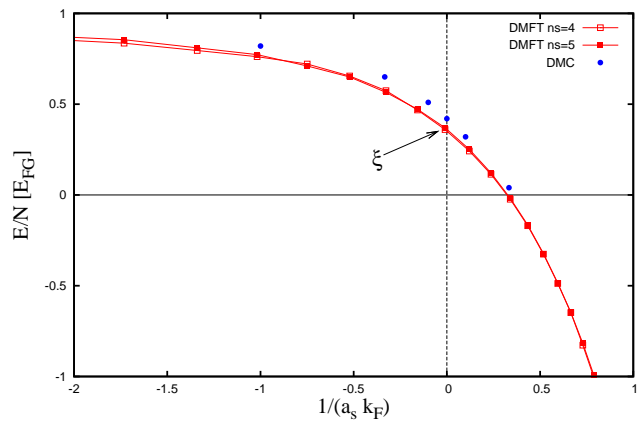}%
\caption{\label{fig:e2n_akf} (Color online)
The $T=0$ energy per particle as a function of the dimensionless parameter
$1/(a_s k_F)$. DMFA at lattice filling of $n=0.1$ with $n_s=4,5$ are denoted
by filled and empty squares. The circle are the DMC of \cite{Carlson03}.}
\end{figure}

{\it Results --} The attractive Hubbard model and the
BCS-BEC cross were originally studied within the DMFT 
by Keller {\it et.} al \cite{Keller01} who have established the phase diagram
of the system at quarter filling $n=1/2$ using a half-ellipse density of
states. Later on this work was followed by \cite{Capone02,Toschi05a,Garg05}
who have 
studied different aspects of the transition in the limit 
$d \longrightarrow \infty$. In \cite{Nir08} we have studied
the continuum limit of the 
metastable Fermi liquid phase in $d=3$. Here we extend this study to the ground
state 
superconducting phase. The properties of a Fermi gas with attractive
interaction where studied by different groups 
using different techniques
\cite{Carlson03,GFMC04,Giorgini04,Bulgac06,Bulgac08}.  
At unitarity, it is customary to present the $T=0$ ground state energy per
particle in the form
$E/N=\xi E_{FG}$ where $E_{FG}=0.6 E_F$ and $E_F$ is the Fermi energy.
The  value $\xi = 0.44 \pm 0.01$ was
calculated by \cite{Carlson03,GFMC04,Giorgini04}, using the 
fixed-node diffusion Monte-Carlo (DMC) method.  
In Fig. \ref{fig:e2n_akf} we present the energy per particle $T=0$ DMFA
results as a function of the interaction strength in comparison with the DMC
results. It can be seen that the DMFA results are in agreement with the
DMC calculations.

\begin{figure}[ht]
\includegraphics[height=6cm]{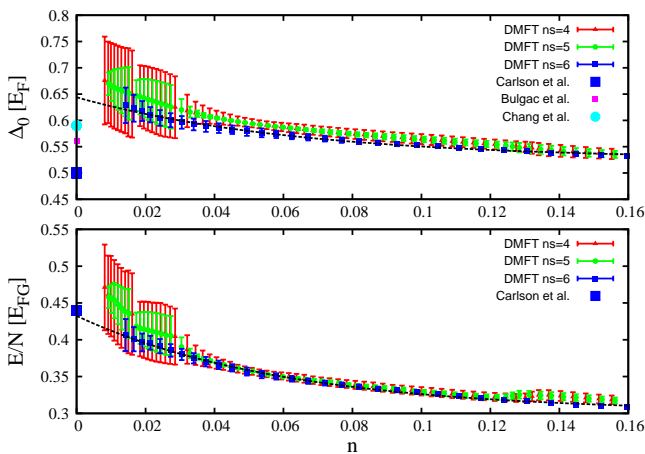}%
\caption{\label{fig:e2n_u} (Color online)
The $T=0$ energy per particle and superconducting gap as a function of the
lattice filling $n$.
The DMFA calculation with $n_s=4$ are shown with triangles, $n_s=5$ with
circles, and $n_s=6$ with squares.
Also shown are the QMC results of \cite{Carlson03,Bulgac08,GFMC04}.}
\end{figure}

Since the continuum corresponds to the limit of vanishing lattice filling
we explore in Fig. \ref{fig:e2n_u} the dependence at unitarity of $E/N$ and
the gap 
function $\Delta_0$ on n. As the lattice filling
approach zero the accuracy of our calculation deteriorates since with $n_s\leq
6$ 
the bath fermions are unable to reproduce the fine details of 
$\hat {\cal G}_0$ needed for such calculation. To demonstrate this point we
attached to each data point an error bar, $\delta_n(E/N)$ to the energy points
and 
$\delta_n \Delta_0$ to the gap points, where
$\delta_n$ is the relative density error discussed above.
From the figure we can see that as expected the error decrease with the number
of auxiliary fields. It can also be seen that the error grows substantially
at the lower most densities.
Extrapolating the $n_s=6$ calculation to the
continuum limit we obtain 
$\xi \approx 0.44 $ and $\Delta_0 \approx 0.64 E_F$. These values
are in close agreement to the Quantum Monte-Carlo (QMC) calculations of
\cite{Carlson03,Bulgac08,Giorgini04}.
 
\begin{figure}
\includegraphics[height=6cm]{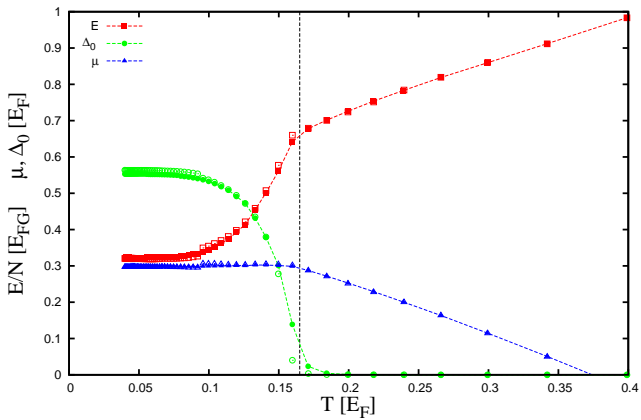}%
\caption{\label{fig:emg_t} (Color online)
The temperature dependence of $E, \mu, \Delta_0$ at unitarity, for lattice
filling $n=0.1$.
The energy per particle $E/N$ is shown with squares, 
the chemical potential $\mu$ is shown with triangles, and the gap function
$\Delta_0$ is shown with circles. Calculations with $n_s=6$ are shown with
filled shapes, calculations with $n_s=5$ with empty ones.  
}
\end{figure}

Turning now to study the behavior of the system at finite temperature 
we present in Fig. \ref{fig:emg_t} the energy per particle, the chemical
potential and the superconducting gap  at unitarity
as a function of temperature for 
lattice filling $n=0.1$.
From the figure one can clearly identify the superconducting phase
transition associated with the vanishing of the superconducting gap 
and the discontinuity in the derivatives of $\mu$ and $E$.
Similar qualitative behavior was already identified by Bulgac {\it et.} al
\cite{Bulgac06}. The critical temperature $T_c\approx 0.16 E_F$ is somewhat
smaller than their result $T_c=0.23 E_F$  but agrees 
with the value $T_c=0.15 E_F $ of Burovski {\it et.} al \cite{Burovski06}.
In the superconducting region, the  gap function behaves approximately
as $\Delta_0(T)=\Delta_0(0)(1-(T/T_c)^{\kappa})$ and the energy 
as $E(T)=N E_{FG}(\xi+\zeta(T/T_c)^{\lambda})$. Comparing the DMFA results
with these formulas we have found that $\kappa \approx 7 $, and $\lambda
\approx 5$. These values are substantially higher than the QMC values
$\kappa=1.5, \lambda=2.5$ found in \cite{Bulgac06}. 
An experimental indication for a large $\lambda$ can be found
in \cite{Kinast05} were the value $\lambda=3.73$ was reported.
for a unitary Fermi gas in a trap.
It is interesting to note that in contrast with the observation of 
Toschi {\it et.} al \cite{Toschi05b} our calculations seems to indicate 
a coincidence between the disappearance of the gap and the discontinuity
of the heat capacity. Thus 
$\Delta_0$ provides no indications for a phase transition 
associated with the closure of a pseudo gap above $T_c$.
On the other hand, in our calculations the gap function $\Delta_0$ doesn't
vanish at $T=T_c$ but, 
like the other thermodynamic
variables, change its behavior and drops to zero exponentially.

{\it Conclusions --} In this work we have applied the DMFA to study 
the properties of a unitary Fermi gas. 
We have found that the predictions of
this theory agree remarkably well with the results of full QMC
simulations. The DMFA energy per particle actually coincide with the 
results of the fixed node Monte-Carlo simulations, whereas the gap function
is somewhat higher than previous estimates.
At finite temperature the DMFA results agree qualitatively with those of
\cite{Bulgac06} although the thermodynamic functions exhibit a stronger
temperature dependence.
These results indicate that the self-energy of a dilute Fermi gas
has only a weak momentum dependence, since by construction
the DMFA equations 
yields the local approximation for the self-energy 
$\hat\Sigma(\bs{k},i\omega_n)\approx \hat\Sigma(i\omega_n)$.
The momentum dependence of $\hat\Sigma(i\omega_n)$ can be investigated by
extending the impurity site from a single node into a cluster of nodes.
The DMFA results can be further refined using a QMC approach at finite
temperature or 
the Lanczos diagonalization method at $T=0$. 
\vspace{2mm}


I wish to thank A. V. Andreev, G. F. Bertsch, A. Bulgac, S. Y. Chang 
and D. Gazit for 
useful discussions and help 
during the preparation of this work.
This work was supported by the Department of Energy Grant
No. DE-FG02-00ER41132. 

\end{document}